\documentclass[11pt]{article}     
\usepackage{osid,graphics,graphicx,axodraw,subfigure}

\newcommand{\bra}[1]{\langle #1 |}
\newcommand{\ket}[1]{| #1 \rangle}
\newcommand{\braket}[2]{\langle #1 \vert #2 \rangle}

\newcommand{\tr}{\mbox{tr}}

\title{A pedagogical overview of quantum discord}
\author{Kavan Modi\footnote{~Previously at Department of Physics, Clarendon Laboratory, University of Oxford, Oxford UK \& Centre for Quantum Technologies, National University of Singapore, Singapore}
\\{\footnotesize\it School of Physics, Monash University, Victoria 3800, Australia\\ \sc{Email:} kavan.modi@monash.edu}}

\begin{document}
\maketitle

\begin{abstract}
Recent measures of nonclassical correlations are motivated by different notions of classicality and operational means. Quantum discord has received a great deal of attention in studies involving quantum computation, metrology, dynamics, many-body physics, and thermodynamics. In this article I show how quantum discord is different from quantum entanglement from a pedagogical point of view. I begin with a pedagogical introduction to quantum entanglement and quantum discord, followed by a historical review of quantum discord. Next, I give a novel definition of quantum discord in terms of any classically extractable information, a approach that is fitting for the current avenues of research. Lastly, I put forth several arguments for why discord is an interesting quantity to study and
why it is of interest to so many researchers in the community. 
\end{abstract}
       
\date{\today}

\section{Introduction}

Quantum systems are correlated in ways inaccessible to classical objects. A distinctive quantum feature of correlations is quantum entanglement~\cite{PhysRev.47.777, Schrodinger:1935eq, PhysRev.48.696, PhysRevA.40.4277}. Entangled states are nonclassical in the sense that they cannot be prepared with the help of \emph{local operations and classical communication} (LOCC)~\cite{horodecki:865}. More recently, another notions of quantum correlations has been proposed, it is called quantum discord~\cite{PhysRevA.71.062307, arXiv:1107.3428, Modi:2011wv}. I show how under the LOCC paradigm both quantum entanglement and quantum discord are meaningful quantities but from a different perspective. 

I begin with a simple explanation of the LOCC paradigm, then show how entanglement is defined and then define quantum discord in the same context. Next, I give a technical discussion on the historical origins of quantum discord and what are the ways in which researchers think about discord at the present. At the end of the article I discuss why quantum discord should be interesting to researchers working in quantum information theory, quantum computation, and foundations of quantum mechanics.

\section{Local operations and classical communication (LOCC)}

Let me begin by introducing the popular characters of quantum information: Alice and Bob. Suppose Alice is in Barcelona and Bob is in Oxford in their respective laboratories with only a telephone line between their labs. They each have one qubit\footnote{A qubit or quantum bit is a quantum system with two possible states: $\ket{0}$ and $\ket{1}$. Being a quantum system a generic qubit state can take any superposition of $\ket{0}$ and $\ket{1}$: $\ket{\psi} = \alpha \ket{0} + \beta \ket{0}$ with $|\alpha|^2 + |\beta|^2 =1$.}, with each qubit in state $\ket{0}_A$ and $\ket{0}_B$. Together I can write the two qubits as $\ket{0}_A \otimes \ket{0}_B = \ket{00}$. 

Now suppose Alice and Bob each have a fair coin. They flip the coin and if only one of them gets `heads' then she/he changes her/his state from $\ket{0}$ to $\ket{\psi}$, and if they both get `heads' then they change the state to $\ket{\phi}$ and $\ket{\chi}$ respectively. The state after the coin toss is
\begin{eqnarray}
\mbox{tails-tails}: \ket{0 0}, \quad \mbox{tails-heads}: \ket{0 \psi}, \nonumber\\
\mbox{heads-tails}: \ket{\psi 0}, \quad \mbox{heads-heads}: \ket{\phi\chi}.\label{exstate}
\end{eqnarray}
Since their action depends on the outcomes of both coins they will need to communicate via the telephone line, this is the {\sl classical communication} part of LOCC. After the communication they can make the appropriate unitary transformation on their respective system, which is the {\sl local operation} part of LOCC\footnote{In general they are not restricted to local unitary operations, rather they are allowed to make arbitrary completely positive maps for local operations.}.

In general they can construct a computer program that has $K$ outcomes. Each outcome occurs with probability $p_k$ and they prepare a corresponding quantum state $\ket{\psi_k \phi_k}$. Let me write the resulting state as an ensemble
\begin{equation}
\{p_k, \, \ket{\alpha_k \beta_k}, \, \ket{k} \}
\end{equation}
or as a density operator
\begin{equation}\label{eqsepgk}
\rho_{AB|k} = \sum_k^K p_k \ket{\alpha_k \beta_k} \bra{\alpha_k \beta_k} \otimes  \ket{k}\bra{k}
\end{equation}
where $\ket{k}$ is a classical flag that stores the information on the outcome of the computer program. In other words I restrict $\braket{k}{k'} = \delta_{kk'}$. This is why I write the subscript $AB|k$ with the latter part said as `given $k$'.

Now suppose Alice and Bob loose the records of the outcomes $k$. In the case of the two coins their state has to be averaged over all possible outcomes of the coin flips
\begin{equation}
\rho_{AB}= \frac{1}{4} \left(\ket{00} \bra{00} + \ket{0\psi} \bra{0\psi}+ \ket{\psi0}\bra{\psi0}+\ket{\phi\chi}\bra{\phi\chi} \right).
\end{equation}
For the general case imagine that the classical register, that retains the information about the value of $k$, is lost. The average state in this case is called a separable state.

{\bf Definition.} \emph{A state is said to be separable correlated if and only if it can be prepared via local operations and classical communication.}
\begin{equation}\label{eqsep}
\rho_{AB}= \sum_k^K p_k \ket{\alpha_k \beta_k} \bra{\alpha_k \beta_k}
 \;\Longleftrightarrow\; \mbox{Separable.}
\end{equation}

\begin{figure}[t]
\subfigure[$\;$ Separable states]{\includegraphics[width=10cm]{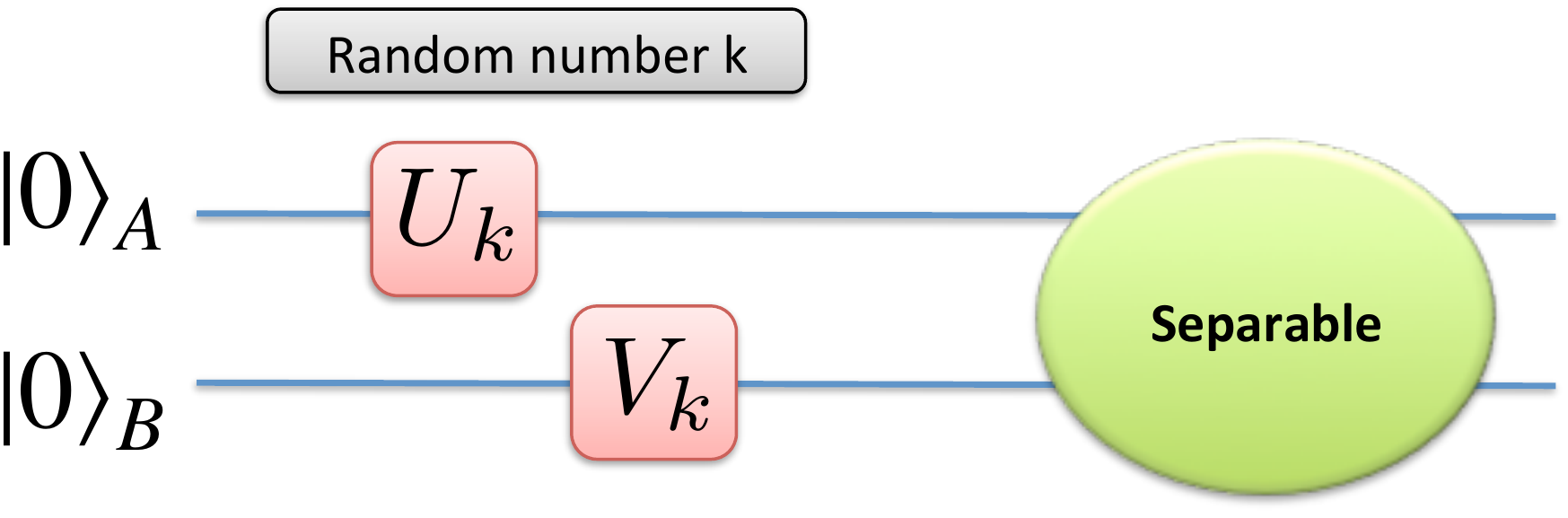}}
\subfigure[$\;$ Entangled states]{\includegraphics[width=10cm]{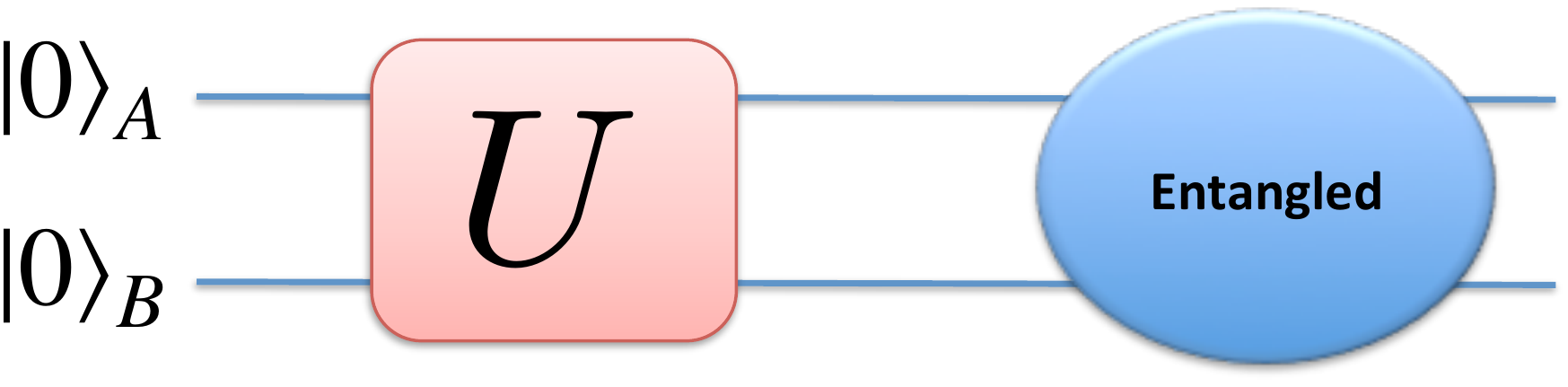}}
\caption{\emph{Preparation of states.}  The Figure (a) shows how local operations and classical communication can be utilised to prepare separable states. A random number $k$ is generated with probability $p_k$, and its value determines the unitary transformation that Alice and Bob will make on their respective systems. If I define $U_k\ket{0} = \ket{\alpha_k}$ and $V_k\ket{0} = \ket{\beta_k}$, then I will generate the state in Eq.~(\ref{eqsepgk}) and Eq.~(\ref{eqsep}). The Figure (b) shows that to prepare entangled states, the two party must interact with each other.
\label{PREP}}
\end{figure}

Now I am ready to introduce quantum entanglement. The procedure detailed in the preceding subsection is called LOCC preparation. Alice and Bob started with a product state and two coins (or a computer program) and by forgetting the outcome of the coin they created a correlated mixed state. An entangled state is a state that cannot be prepared in this manner. In mathematical sense, an entangled state is the one that cannot be written as a convex mixture of product states. An entangled state is prepared by a genuine quantum interaction between Alice and Bob, i.e., a {\emph global} unitary transformation between them. It is generally very difficult to characterise entanglement and formally entanglement is defined as a non-separable state.

{\bf Definition.} \emph{A state is said to be entangled correlated if and only if it {\bf cannot} be prepared via local operations and classical communication.}
\begin{equation}
\rho_{AB} \ne \sum_k^K p_k \ket{\alpha_k \beta_k} \bra{\alpha_k \beta_k} \;\Longleftrightarrow\; \mbox{Entangled.}
\end{equation}

The two concepts of separable states and entangled states are depicted in Fig.~\ref{PREP}.

\section{Classically correlated states}
A few of natural questions arise: {\sl Is the concept of quantum correlations the same as entanglement?} {\sl Are separable states same as classically correlated states?} Add another question to this list: {\sl What is the state of a classically correlated system?} 

In classical information theory one only needs to worry about bits, taking values 0 or 1. The state of a quantum object is in general described by a density operator $\rho_{AB}$, while the state of a (correlated) classical system is described by a joint probability distribution $P_{ab}$. Although, I can write the state of a classically correlated system as density operator as $\rho_{AB} = \sum_{ab} p_{ab} \ket{ab} \bra{ab}$, where $\{\ket{ab}\}$ forms an orthonormal basis.

As an example consider the state 
\begin{equation}\label{maxclcor}
\rho_{AB} = \frac{1}{2} \left(\ket{00}\bra{00} + \ket{11}\bra{11} \right)
\end{equation}
is a classically correlated state. However, I can also construct a correlated classical state in a different basis:
\begin{eqnarray}
\rho_{AB} &=& \frac{1}{2} \left(\ket{x_0 x_0}\bra{x_0 x_0} + \ket{x_1 x_1}\bra{x_1 x_1} \right) \quad \mbox{or}\\
\rho_{AB} &=& \frac{1}{2} \left(\ket{x_0 y_0}\bra{x_0 y_0} + \ket{x_1 y_1} \bra{x_1y_1} \right)
\end{eqnarray}
with $\braket{x_0}{x_1}= 0$ and $\braket{y_0}{y_1}= 0$.

A state is classically correlated as long as the total state is diagonal in an orthonormal product basis. One reason for calling such a state classically correlated is that it can be measured and determined without altering it. In other words, if I know that the state is diagonal in basis $\ket{ab}$, then I can determine the values of $p_{ab}$ without disturbing the state. I will adopt this to be formal definition of a classically correlated state. This definition then gives me the set of correlated nonclassical states by excluding the classically correlated state from the set of all states.

{\bf Definition.} \emph{A state is said to be classically correlated if and only if it can be fully determined without disturbing it with the aid of local measurements and classical communication.}
\begin{equation}
\rho_{AB} = \sum_{ab} \ket{ab} \bra{ab} \rho_{AB} \ket{ab} \bra{ab}
=\sum_{ab} p_{ab} \ket{ab} \bra{ab}  \;\Longleftrightarrow\; \mbox{classically correlated},
\end{equation}
where $\{\ket{ab}\}$ forms an orthonormal basis, i.e., $\braket{ab}{a'b'}=\delta_{aa'} \delta_{bb'}$.

\subsection{Quantum discord}
I am now ready to define the spirit of quantum discord. The following definition departs from the historical definition, however is motivated in similar spirit as the definition of quantum entanglement.
We simply say that a bipartite state has  quantum discord when it is not a classically correlated state\footnote{In much of the literature there is a distinction made between symmetric quantum discord and asymmetric quantum discord. Here I only work with the symmetric and asymmetric versions of quantum discord interchangeably.}.

{\bf Definition.} \emph{A state is said to be discordant if and only if it {\bf cannot} be fully determined without disturbing it with the aid of local measurements and classical communication.}
\begin{equation}
\rho_{AB} \ne \sum_{ab} \ket{ab} \bra{ab} \rho_{AB} \ket{ab} \bra{ab}
=\sum_{ab} p_{ab} \ket{ab} \bra{ab} \;\Longleftrightarrow\; \mbox{discordant},
\end{equation}
where $\{\ket{ab}\}$ forms an orthonormal basis, i.e., $\braket{ab}{a'b'}=\delta_{aa'} \delta_{bb'}$.

Note that the difference between a separable state and an entangled state has to with how each state is prepared. While the difference between a classically correlated state and a discordant state has to with whether the details of the state can be measured without disturbing the state. We will return to this issue at a later point.

\begin{figure}[t]
\includegraphics[width=10cm]{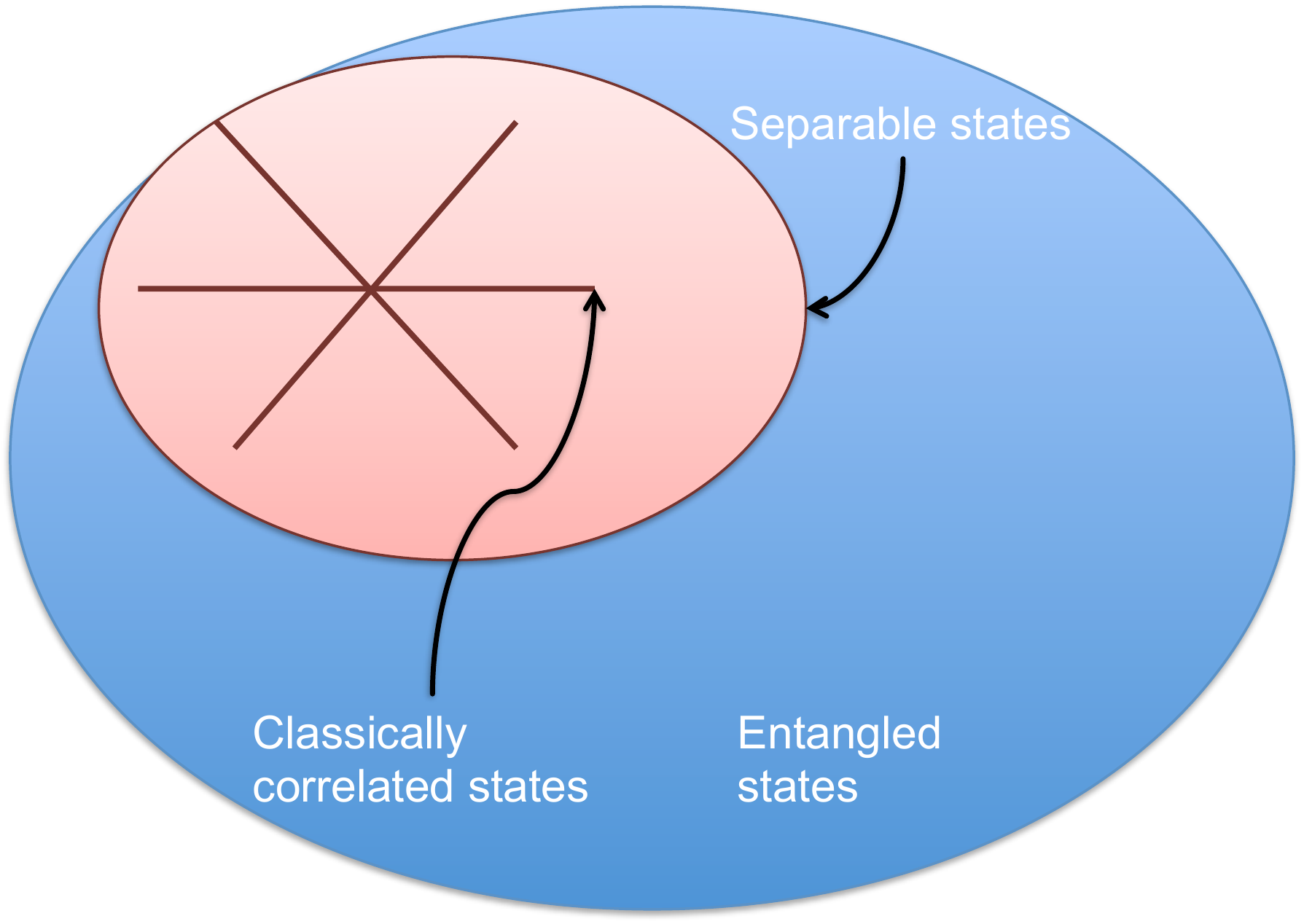}
\caption{\emph{States.}  The large ellipse represents the set of all states with the set of separable states in the smaller ellipse. The lines represent the set of classically correlated states (in different basis). The point where the lines meet is the maximally mixed state, which is classical in any basis. The other end of the black line then should be a pure state in product form. All states other than the ones on the lines have quantum discord. The entangled states, discordant states, and classically correlated states do not form a convex set. That is, mixing two entangled (discordant) states can lead to a separable (classical) state. Only the separable states do form a convex set. Indeed this is one troubling aspect of quantum discord; locally mixing classically correlated states can lead quantum discord. In other words, one can use LOCC to prepare discordant state (as mentioned in text).
\label{CORRELATIONS}}
\end{figure}

The set of classically correlated states are a subset of separable states. According to the definition above some separable states have quantum discord. This means that quantum discord can be generated by LOCC. On the other hand all entangled states are also discordant. The most important point is that discordant states are not like classically correlated states. In that sense it is a weaker criteria for quantumness than entanglement. In Fig.~\ref{CORRELATIONS} I depict the four variety of states and in caption discuss some of their features. In the next section I will put forth arguments for and against quantum discord whether it can capture some features of quantumness even when there is no entanglement present. But first let me give a historical introduction of quantum discord. Then I will redefine quantum discord in a technical manner that is more fitting to the current avenues of research.

\section{Historical origin of quantum discord}

The story of quantumness of correlations beyond-entanglement begins with the non-uniqueness of quantum conditional entropy. Let $P_{ab}=\{p_{ab}\}$ be a joint classical probability distribution. The marginal distributions can be attained by summing over one of the indices: $P_a = \{\sum_b p_{ab} \}$ and $P_b = \{\sum_a p_{ab} \}$. Additionally a conditional distribution is defined as $P_{b|a} = \{p_{ab}/p_a\}$. A way of quantifying the uncertainty of a probability distribution is by its entropy: 
\begin{equation}\label{sent}
H(ab) = H(P_{ab}) =-\sum_{ab} p_{ab} \log(p_{ab}).
\end{equation}

\begin{figure}[t]
\resizebox{7.67 cm}{5.87 cm}{\includegraphics{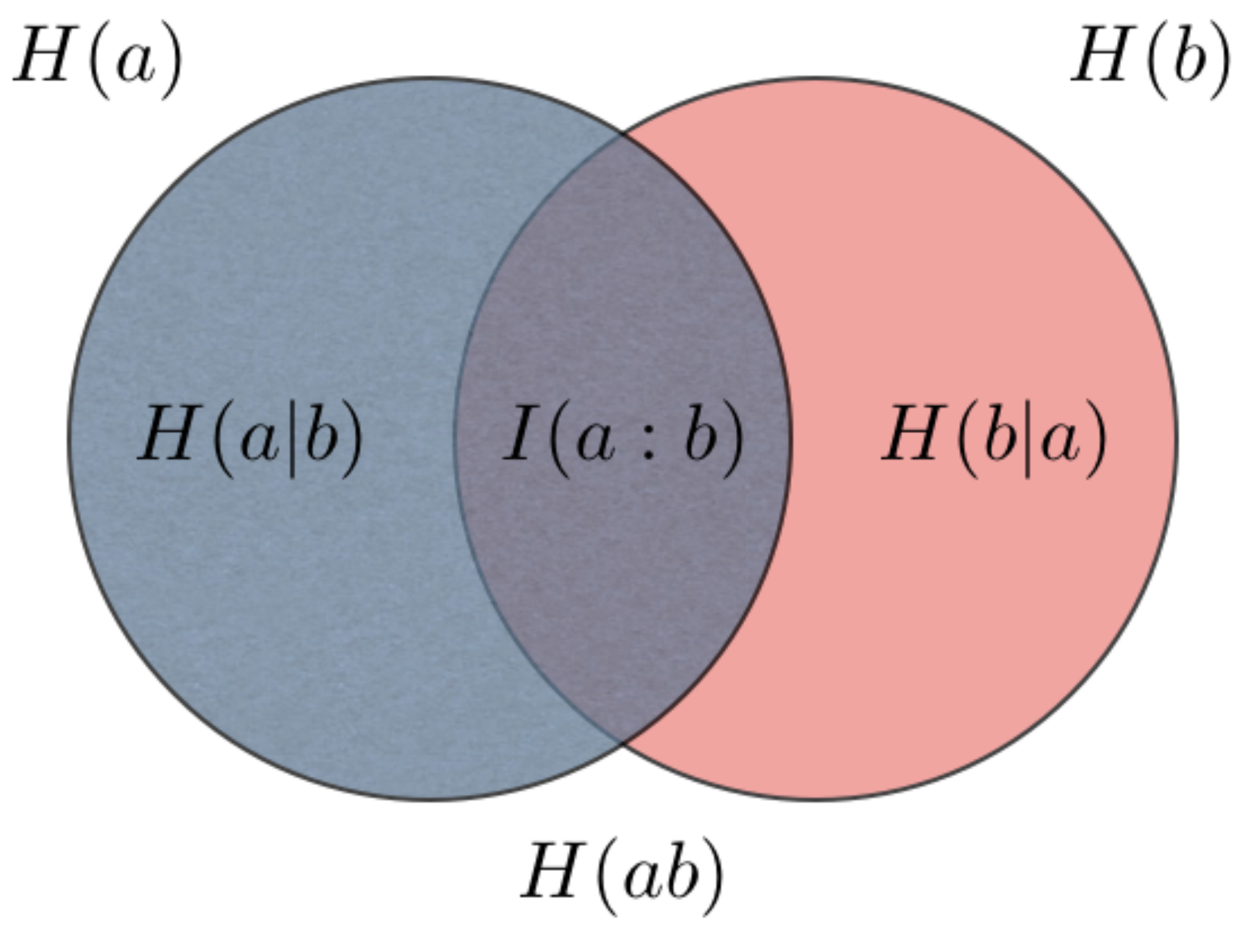}}
\caption{\emph{Conditional entropy.} The Venn diagram shows the joint entropy $H(ab)$, marginal entropies $H(a)$ and $H(b)$, conditional entropies, $H(a|b)$ and $H(b|a)$, and mutual information $I(a:b)$ for a joint classical probability distribution for (correlated) random variables $a$ and $b$.\label{condent}}
\end{figure}

In classical probability theory one may define conditional entropy as
\begin{equation}
H(b|a) = H(ab)-H(a).\label{clce1}
\end{equation}
It is the measure ignorance of $b$ given some knowledge of state of $a$. Fig.~\ref{condent} depicts this relationship in a graphical manner. Another way to express the conditional entropy is as the lack of knowledge of the value of $b$ when the state of $a$ is known to be in the $a$th state, weighted by the probability for $a$th outcome as
\begin{eqnarray}
H(b|a) &=& \sum_{a} p_a H(P_{b|a})\label{clce2}\\
&=&-\sum_a p_a \sum_{b} \frac{p_{ab}}{p_a} \log\left( \frac{p_{ab}}{p_a} \right) \\
&=&-\sum_{ab} p_{ab} \log (p_{ab}) + \sum_{ab} p_{ab} \log (p_a) \\
&=& H(ab) - H(a).
\end{eqnarray}

The classical-equivalent of Eqs.~(\ref{clce1}) and~(\ref{clce2}) give rise to quantumness of correlations and in specific quantum discord~\cite{datta08}. This is due to the fact that these two equations are not the same in quantum theory. While the first simply takes the difference in the joint ignorance and the ignorance of $a$, the second equation depends on specific outcomes of $a$, which requires a measurement. However, measurements in quantum theory are basis dependent and change the state of the system. 

\subsection{Quantum conditional entropy}

In generalising the classical concepts above to quantum I replaced the classical-probability distributions with density operators and Shannon's with von Neumann's entropy:
\begin{eqnarray}\label{vnent}
S(AB) = S(\rho_{AB}) &=& -\tr[\rho_{AB} \log(\rho_{AB})] \\
&=& -\sum_{ab} \lambda_{ab} \log(\lambda_{ab}).
\end{eqnarray}
where $\lambda_{ab}$ are the eigenvalues of $\rho_{AB}$.

Now, how can I deal with conditional entropy then? Clearly there are at least two options\footnote{A different approach to conditional entropy is taken in~\cite{PhysRevLett.79.5194, PhysRevA.60.893}, where a quantum \emph{conditional amplitude} (analogous to classical conditional probability) is defined such that it satisfies Eq.~(\ref{qce1}). I only mean to suggest that the two approaches above are not the only options available. Different approaches give different distinctions of quantum theory from the classical theory. And in someway different notions of quantumness.}, namely Eqs.~(\ref{clce1}) and~(\ref{clce2}). Let me deal with Eq.~(\ref{clce1}) first and define quantum conditional entropy as
\begin{equation}\label{qce1}
H(b|a) \rightarrow S(B|A)=S(AB)-S(A).
\end{equation}

This is a well known quantity in quantum information theory~\cite{Schumacher} and negative of this quantity is known as coherent information. However, this is a troubling quantity as it can be negative for entangled states. Imagine the quantum state $\frac{1}{\sqrt{2}}(\ket{00} + \ket{11}).$ This is a pure state and so it has zero entropy, while its marginal has maximal entropy. For a long time there was no way to interpret the negativity~\cite{merging,cooling}. This is in stark contrast with the classical conditional entropy, which has a clear interpretation and is always positive.

On the other hand, I can define the quantum version of Eq.~(\ref{clce2}) by making measurements on party $A$. To put in the details, the joint state $\rho_{AB}$ is measured by $A$ giving $a$th outcome:
\begin{equation}
\rho_{AB}\rightarrow \sum_a \Pi_a \rho_{AB} \Pi_a = \sum_a p_a \ket{a}\bra{a} \otimes \rho_{B|a},
\end{equation}
where $\Pi_a$ are rank-one \emph{positive operator values measures} (POVM), $\ket{a}$ are classical flags on measuring apparatus indicating the measurement outcome, $p_a = \mbox{Tr}[\Pi_a \rho_{AB}]$ is probability of $a$th outcome, $\rho_{B|a} = \mbox{Tr}_A[\Pi_a\rho_{AB}]/p_a$. The conditional entropy of $B$ is then clearly defined as
\begin{equation}\label{qce2}
H(b|a) \rightarrow S(B|\Pi_A)=\sum_a p_a S(\rho_{B|a}).
\end{equation}
This definition of conditional entropy is always positive. The obvious problem with this definition is that the state $\rho_{AB}$ changes after the measurement. Also note that this quantity is not symmetric under party swap.

Clearly the two definitions of conditional entropies above are different in quantum theory. The first one suffers from negativity and the second one needs `classicalisation' of a quantum state. 

\subsection{The original discord}

Let me now derive quantum discord and relate it to the preceding section. I start with the concept of mutual information. Both in quantum and classical case it is defined as
\begin{eqnarray}
I(a:b) &=&H(a)+H(b)-H(ab) \quad \mbox{classical}\\
I(A:B) &=& S(A) + S(B) - S(AB) \quad \mbox{quantum}.
\end{eqnarray}
For the classical using Eq.~(\ref{clce2}) it can also be defined as
\begin{equation}
J(b|a)=H(b)-\sum_{a} p_a H(b|a).
\end{equation}
Mutual information is a measure for the information stored in the joint state, i.e., in correlations (see Fig.~\ref{condent} for a graphical depiction). For instance consider the state in Eq.~(\ref{maxclcor}).

The two classical-equivalent mutual information above are not the same in quantum theory. The mutual information one can utilise via measuring $A$ first and then $B$ is different from measuring $AB$ together. This is precisely what was noted by several authors in near temporal proximity~\cite{henderson01a, PhysRevLett.88.017901, PhysRevLett.89.180402}. Henderson and Vedral~\cite{henderson01a} called  $J(B|\Pi_A)$ classical correlations because it is the information gained by measuring the system. Ollivier and Zurek gave the difference between $I(A:B)$ and $J(B|\Pi_A)$ its now famous name \emph{quantum discord}:
\begin{eqnarray}
\delta(B|A)&=&I(A:B)-J(B|\Pi_A)\\
&=&S(B|\Pi_A)-S(B|A).
\end{eqnarray}
Working out the details one finds that quantum discord is simply the difference between two definitions of conditional entropy. See~\cite{arXiv:1104.1520} for other ways to construct measures of quantum discord and~\cite{arXiv:1108.3649} for a set of reasonable criteria one should have in a measure of quantum discord.

\subsection{Extremisation conditions}

In order for discord to be not a function of the choice of the performed measurement Henderson and Vedral~\cite{henderson01a} advocated for a maximisation of classical information from correlations over all POVMs. This leads to the most popular formula for quantum discord
\begin{equation}
\delta(B|A)=\min_{\Pi_A}[I(A:B)-J(B|\Pi_A)].
\end{equation}
Since conditional entropy in Eq.~(\ref{qce2}) is asymmetric under party swap, quantum discord is also asymmetric under party swap. 

Oppenheim et al.~\cite{PhysRevLett.89.180402} argued for maximisation of classical correlations over the whole set of LOCC protocols. If I denote the final state of Alice and Bob as $\rho_{A'}$ and $\rho_{B'}$ after the LOCC protocol respectively. Then the deficit is defined as:
\begin{equation}
\Delta(A:B)=\min _{\Lambda_{AB}}[S(A')+S(B')]-S(AB),
\end{equation}
where $\Lambda_{AB}$ are the set of generalised operations implementable by LOCC. This quantity is called quantum deficit. The meaning of quantum deficit is the smallest amount of information contained in correlation that cannot be attained by LOCC. 

Entangled states certainly have finite deficit, but so do some separable states. Let me give an example of a bipartite state that has finite discord but zero two-way deficit:
\begin{equation}
\rho_{ABC} = \frac{1}{2} \left(\rho_{Ab} \otimes \ket{0}\bra{0} + \rho_{aB} \otimes \ket{1}\bra{1} \right),\label{deficitstate}
\end{equation}
where
\begin{eqnarray}
\rho_{Ab} &=& \frac{1}{2} \left(\ket{00}\bra{00} + \ket{+1}\bra{+1}  \right)\\
\rho_{aB} &=& \frac{1}{2} \left(\ket{00}\bra{00} + \ket{1+}\bra{1+}  \right).
\end{eqnarray}
The state in Eq.~(\ref{deficitstate}) has finite discord. However, if party $C$ reveals the value of his qubit's state, then $A$ and $B$ can devise a LOCC protocol to determine their state exactly. In other words, the information about the state gained by LOCC is the same as information gained by a global measurement, i.e., the quantum deficit is zero.

\subsection{Discord reformulated}

The guiding principle of quantum discord can be stated as the following. {\sl The information that one can acquire from the correlations in a bipartite quantum state using the LOCC prescription is less than the total information stored in the correlations if and only if the bipartite state is discordant.} In other words, the information that one can acquire from the correlations without disturbing the system is considered classical. Quantum discord is the complement of the preceding statement and quantifies the disturbance caused by a measurement.

With this guiding principle I can reformulate the notion of classical acquisition of information (the complement of quantum discord) following the arguments of Brodutch~\cite{aharon}. Let Alice and Bob make any LOCC operation they like. Let $\Lambda_{AB} \in \mbox{LOCC}$ and 
\begin{equation}
\Lambda_{AB}(\rho_{AB}) \to \rho'_{AB}. 
\end{equation}
I will call the state of Alice and Bob `partially classical' if there exists a non-trivial $\Lambda_{AB}$ leading to real information that does not alter the mutual information in the state:
\begin{equation}
I(A:B) = I(A':B') \quad \Longleftrightarrow \quad \mbox{partially classical}.
\end{equation}
The technical arguments for the above definition rely on a beautiful theorem due to Petz. The theorem shows that any $\Lambda_{AB}$ that does not change the mutual information there exists a $\Gamma_{AB}$ such that 
\begin{equation}
\Lambda_{AB} (\rho_{AB})  \to \rho'_{AB} \quad \mbox{and} \quad 
\Gamma_{AB} (\rho'_{AB} ) \to \rho_{AB},
\end{equation}
i.e., the operations are reversible. See~\cite{aharon} and references therein for details. Lastly, note that if there does not exist a measurement that gains some information about either $A$ or $B$ without disturbing the state then quantum deficit is finite.

Let me give an example to describe the new notion of classicality. Suppose Alice and Bob have one of two states: 
\begin{equation}
\frac{1}{\sqrt{2}} (\ket{00}+\ket{11}) \quad \mbox{or} \quad \frac{1}{\sqrt{2}} (\ket{2+}+\ket{3-}). 
\end{equation}
Now Alice can make a measurement on her system with projectors 
\begin{equation}
P_{01}= \ket{0}\bra{0} + \ket{1}\bra{1} \quad \mbox{and} \quad P_{23}= \ket{2}\bra{2} + \ket{3}\bra{3}. 
\end{equation}
She will observe one of two outcomes and she can communicate that to Bob, which will enable them to share an entangled state at the end of the day. Clearly neither state of $AB$ is classically correlated since there is entanglement. However, the measurement does not alter the state and therefore the extracted information is classical. Note that there is no measurement Bob can make that will reveal to him which state he shares with Alice. This is because states $\{\ket{0}, \, \ket{1}\}$ and state $\{\ket{+}, \, \ket{-}\}$ are overlapped, i.e., cannot be measured simultaneously. The key difference between this notion of classicality and discord is that, here I simply mean that there exists some information in the state that is retrievable without altering the state, while the resulting state maybe entangled.

\section{Why discord is worth studying}

Classically correlated states are described by a joint probability distribution. A classical computer of $n$ bits has a state at any point in time that is fully described by a joint probability distribution of $n$ bits. Such a state has no discord at any point in the computation. This then assures that all classical computers can operate without any discord. Now I ask the converse question: {\sl Does the presence of discord yields some quantum enhancement?} In some sense this is the most general question one can pose, with subset to this problem being quantum enhancement in metrology, cryptography, and a whole set of physical problems.

\subsection{Copying correlated quantum states}

Arguably the first major result of quantum information theory is the no-cloning theorem~\cite{Nature.299.802}. The theorem says that unknown quantum state cannot be copied, which enables the field of quantum cryptography. Now suppose Alice and Bob share a quantum state that they wish to copy by some LOCC strategy. This protocol is called local broadcasting and it is shown to be possible if and only if the state is classical~\cite{arXiv:0707.0848, LettMathPhys.92.143}. In some sense the result is straightforward application of the fact that measuring a quantum state necessarily disturbs it. Only when a quantum state can be interrogated without alteration it is possible to clone.

\subsection{Quantum-classically correlated states}

One of the first utility of quantum information was for cryptography in the protocol called BB84~\cite{bb84} (developed by Bennett and Brassard in 1984). As noted above quantum cryptography is an application that makes use of the no-cloning theorem. In this protocol Alice prepares one of four states $\{\ket{0}, \, \ket{1}, \, \ket{+}, \, \ket{-} \}$ and sends it to Bob. The average state has the form
\begin{equation}
\rho_{\mbox{BB84}}= \frac{1}{4}(\ket{00}\bra{00} + \ket{11}\bra{11} + \ket{0+}\bra{0+} + \ket{1-}\bra{1-}),
\end{equation}
a equal mixture of two maximally classically correlated states in two different basis. This state has discord as measured by Bob.

He measures the state in either $\{\ket{0}, \, \ket{1} \}$ basis or $\{ \ket{+}, \, \ket{-} \}$ basis. His outcomes are completely random looking. They do this many times and at the end of the day Alice and Bob publicly announce\footnote{The announcement could be made in a news paper back in 1984 or a in a blog in 2013.} the basis of preparation and measurement respectively. When the basis match they know their results were exactly the same, i.e., perfectly correlated. When basis do not match their results should be randomly correlated. In the latter case, they can announce these results as well on the blog and check if there are any correlations between the preparations and measurements. If correlations exists then they can suspect that someone may have been meddling with the states on their way from Alice to Bob. If they are satisfied with the randomness in correlations of the latter case then they can use the correlated bit string of the former case as a one-time pad, the fundamental object of cryptography.

Now note that there is no entanglement here and it seems that only one-way discord is enough for this application. States with one way discord are often called classical-quantum states and have been identified as resources in other similar applications: quantum locking of classical correlations~\cite{arXiv:1105.2768} and blind quantum computation~\cite{bqc}. The latter makes use of BB84 like cryptography and measurement based quantum computation to ensure that Alice can get Bob to run her program on his quantum computer without him knowing anything about her code, a task that is classically impossible. See~\cite{arXiv:1309.2446} for a discussion on the role of quantum discord in cryptography.

\subsection{Quantum computing and metrology}

Suppose that I have a quantum computer of $n$ qubits such that at every point in the computation its state is classically correlated (in some basis). Such a computation is called a {\sl concordant computation}. A question that still remains unanswered is whether concordant computation can be simulated by a classical computer. In other words, are there problems that can be solved by a concordant computation that cannot be solved on a classical computer? The next question is natural: {\sl For there to be a quantum enhancement in a computation what type of correlations must be present?} Our choices are: concordant classical, discord, and entanglement. Again the answer remains unknown. 

There has been some progress in showing that concordant computation can be classically simulated under some constraints~\cite{arXiv:1006.4402}. While others have put forth arguments for why discord maybe enough to perform some sort of a quantum computation~\cite{arXiv:1109.5549}. Quantum metrology or parameter estimation is an application of quantum computing. Here too entangled states are known to give the optimal results within certain constraints~\cite{PhysRevLett.96.010401}. Here numerical results suggest that quantum enhancement in metrology is present even when entanglement is absent~\cite{arXiv:1003.1174}. The difficulty in all of the problems listed here lies with dealing with the vastness of classical probability distribution in concourse with vastness of the Hilbert space.

Both the arguments for and against discord have merits as well as limitations. Researchers know how a classical computer works and understand how a quantum computer works. The former utilises classically correlated states, while the latter works with pure entangled states. What researchers do not understand is the intermediate situation. In the middle, where a messy mixture of quantum and classical correlations live together, is difficult to work with in theory and unavoidable experimentally. 

This is what quantum discord attempts to characterise, the messy mesoscopic world. This question has practical importance as its answer maybe crucial in operating a real quantum computer. It has foundational importance, as it draws a boundary between quantum and classical worlds.

\subsection{Decoding correlated states}

\begin{figure}[t]
\subfigure[$\;$ Classically correlated states] {\includegraphics[width=10cm]{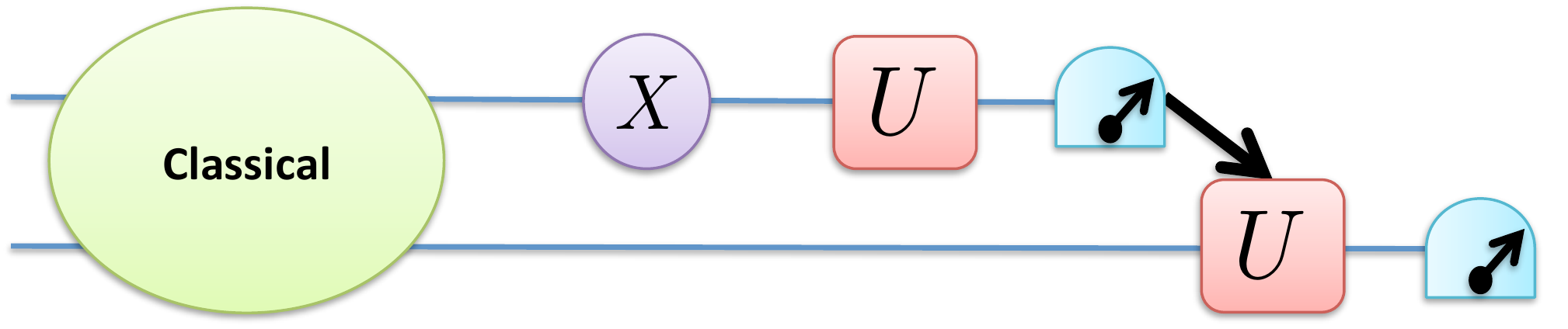}}
\subfigure[$\;$ Discordant states] {\includegraphics[width=10cm]{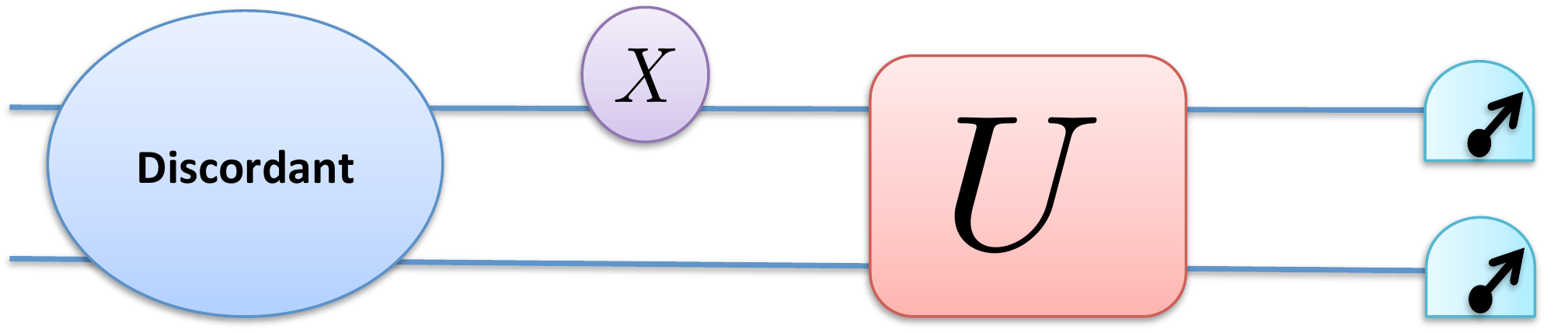}}
\caption{\emph{Decoding states.}  Alice and Bob start with a shared bipartite state. Alice then encodes $X = \{x_k\}$ on her subsystem using local unitary transformations $W_k$ with probability $p_k$. Bob is challenged to determine Alice's choice of $x_k$. Figure (a) shows how an LOCC measurement scheme can be utilised to determine the value of $x_k$ when the initial state is a classically correlated state. The Figure (b) shows that for the best determination of the value of $x_k$ with an initially discordant state Bob must make an entangling measurement. This is the same as making an entangling unitary transformation followed by local measurements. Note that the difference between this Figure and Fig.~\ref{PREP}, where LOCC is utilised to prepare separable states and interaction is necessary for preparing entangled states. For decoding entangling unitary transformation are deemed to be necessary for even discordant states.
\label{DEC}}
\end{figure}

Lastly, let me discuss one last application of discord which contrasts the definition of entanglement. Let me go back to how entangled states are prepared. For any two systems that are entangled they must have had a quantum interaction in the past (see Fig.~\ref{PREP}). They must have exchanged energy and share some quantum information.

Now suppose Alice and Bob share a bipartite quantum state. Alice then encodes a classical variable $X = \{x_k\}$ on her subsystem using local unitary transformations $W_k$ with probability $p_k$. Now Bob is asked to determine the value of $x_k$ by either making an entangling measurement on both parts or performing LOCC operations. In~\cite{arXiv:1203.0011} it is shown for states when the initial state is a classically correlated state Bob is able to reach maximum efficiency in predicting $X$ with just LOCC. While for discordant states he has to utilises entangling operations to reach the maximum efficiency in prediction the value of $X$. This protocol yield one operational interpretation of discord.

Now, compare the statement of preparation of bipartite states to the statement of decoding of a bipartite state. In order to prepare an entangled state with need a global unitary transformation $U$ that interacts the systems of Alice and Bob (see Fig.~\ref{PREP}). While in the decoding protocol (as defined here) a global unitary transformation $U$ is needed to attain the best results for discordant states (see Fig.~\ref{DEC}). The implication is that creating entanglement requires similar resources as decoding discordant states.

\section{Conclusions}

I have given a pedagogical review of ideas behind quantum correlations beyond entanglement. I have defined classically correlated states and discordant states in the spirit of separable states and entangled states. After this I described the historical arguments that led to quantum discord. Next, using the arguments of Brodutch I generalised the concept of quantum discord. Finally, I tried to give several arguments that make quantum discord interesting for a few applications.

\section*{Acknowledgments}
I am grateful for the financial support by the John Templeton Foundation, National Research Foundation, and the Ministry of Education of Singapore. I thank Eduardo Mascarenhas for giving many helpful comments that improved the presentation and readability of this manuscript.

\bibliographystyle{is-unsrt}
\bibliography{modi-discord.bib}

\end{document}